\documentclass{elsarticle}

\usepackage{hyperref}
\usepackage{bm}

\journal{Annals of Physics}

\usepackage{graphicx}  % needed for figures
\usepackage{dcolumn}   % needed for some tables
\usepackage{bm}        % for math
\usepackage{amssymb}   % for math
\usepackage{amsmath}

\usepackage{cancel}
\usepackage{empheq}
\usepackage{graphicx,color}

\def\d{\mathrm{d}}

\begin{document}

\begin{frontmatter}

\title{Boundaries without boundaries}

\author{Paolo Facchi$^{1,2}$, Giancarlo Garnero$^{1,2}$, Giuseppe Marmo$^{3,4}$, Joseph Samuel$^{5}$, Supurna Sinha$^{5}$}

\address{$^{1}$Dipartimento di Fisica and MECENAS, Universit\`a di Bari, I-70126  Bari, Italy}
\address{$^{2}$INFN, Sezione di Bari, I-70126 Bari, Italy}
\address{$^{3}$Dipartimento di Scienze Fisiche and MECENAS, Universit\`a di Napoli ``Federico II", I-80126  Napoli, Italy}
\address{$^{4}$INFN, Sezione di Napoli, I-80126  Napoli, Italy}
\address{$^{5}$Raman Research Institute, 560080 Bangalore, India}

\begin{abstract}

Starting with a quantum particle on a closed manifold without boundary, we consider the process of generating boundaries by modding out by a group action with fixed points, and we  study the emergent quantum dynamics on the quotient manifold.

As an illustrative example, we consider a free nonrelativistic quantum particle on the circle and generate the interval via parity reduction. 
A free particle with Neumann and Dirichlet boundary conditions on the interval is obtained, and, by changing the metric near the boundary, Robin boundary conditions can also be accommodated. We also indicate a possible method of generating non-local boundary conditions. 
 
Then, we explore an alternative generation mechanism which makes use of a folding procedure and is applicable to a generic Hamiltonian through the emergence of an ancillary spin degree of freedom.
\end{abstract}


\end{frontmatter}

\section{Introduction}

Boundary conditions emerge as a model of the interaction of a confined physical system with its boundary. In this paper we will be interested in  applications  to quantum systems. Quantum boundary conditions can be effectively used to describe different quantum situations ranging from topology change in quantum gravity~\cite{reviewbal,wil}, to the Casimir effect~\cite{casimir} in quantum field theory and the quantum Hall effect in condensed matter physics~\cite{morandi}. For general reviews see~\cite{asorey1,asorey2,bangalectures}

From the mathematical point of view quantum boundary conditions emerge as a parametrization of the self-adjoint extensions of a differential operator~\cite{grubb}. It is well known, indeed, from the basic principles of quantum mechanics, that physical  observables correspond to self-adjoint operators~\cite{vneumann}. 

A paradigmatic example is a free nonrelativistic quantum particle in a cavity~$\Omega$, an open bounded set of $\mathbb{R}^n$, whose kinetic energy is described by the Laplace operator:
\begin{equation}
\label{eq:ham1}
H=-\frac{\hbar^2}{2m}\Delta, \quad\quad D(H)= C^\infty_c(\Omega),
\end{equation}
where $m$ is the mass of the particle, $\hbar$ is the Planck constant, and the operator domain $D(H)$ is the space of smooth functions compactly supported in $\Omega$. This  operator is symmetric but not self-adjoint, and admits infinitely many self-adjoint extensions as provided by von Neumann's theory of self-adjoint extensions~\cite{reed}.

There has been an increasing interest in classifying the self-adjoint extensions of elliptic operators in terms of boundary conditions. In particular, it was proved~\cite{asorey1,lledo,extensions} that the set of the self-adjoint extensions of the Laplace operator on a manifold with boundary is in one-to-one correspondence with the set of unitary operators on the boundary. 
The situation can be easily specialized for the one dimensional case. 

Consider the interval $\Omega=(0,\pi)$ and let $L^2(\Omega)$ be the  Hilbert space of square integrable functions on $\Omega$.  The Hamiltonian~\eqref{eq:ham1} reads
\begin{equation}
\label{eq:ham}
H=\frac{p^2}{2m}=-\frac{\hbar^2}{2 m} \frac{{\d}^2}{\d x^2}.
\end{equation}

As already stressed, this operator is not self-adjoint, but admits infinitely many self-adjoint extensions, each of which is parametrized by a two dimensional unitary matrix~\cite{asorey1}. 
Well-known boundary conditions are Dirichlet:
\begin{equation}
\psi(0)=0, \qquad \psi(\pi)=0,
\end{equation}
and Neumann:
\begin{equation}
\psi'(0)=0, \qquad \psi'(\pi)=0,
\end{equation}
where $\psi'= \d \psi/\d x$.
These are examples of local boundary conditions, which do not mix the values at the endpoints of the interval.
The most general local boundary conditions are given by Robin:

\begin{equation}
\label{eq:robin}
\psi'(0)=\mu_0\psi(0)\qquad \psi'(\pi)=-\mu_\pi\psi(\pi), \qquad \mu_0,\mu_\pi\in\mathbb{R}.
\end{equation}
Notice that for $\mu =0$ one recovers Neumann, while for $\mu \to \infty$ one gets Dirichlet. In general, Robin boundary conditions mix the values of the function $\psi$ with  that of its derivative $\psi'$ at the boundary points $x=0$ and $x=\pi$.

A family of non-local boundary conditions is provided by
\begin{equation}
\psi(0)=e^{i\alpha}\psi(\pi),\qquad \psi'(0)=e^{i\alpha}\psi'(\pi),\qquad\qquad \alpha\in\mathbb{R},
\end{equation}
which are known as twisted (or pseudo-) periodic boundary conditions. As a particular case, one recovers periodic and anti-periodic boundary conditions for $\alpha=0$ and $\alpha=\pi$, respectively.

Several problems can be studied with the above technology ranging from one dimensional systems with time-dependent boundary conditions~\cite{compositionlaw, geomphase, compositionlaw1} to quantized fields~\cite{cas1, cas2, cas3}.

The central question of interest in this paper is the following:
\emph{Can one generate boundary conditions starting from a quantum system on a manifold without boundaries?}

In the first part of the paper we will analyze what kind of boundary conditions can emerge via a symmetry reduction procedure starting from a manifold without boundary. Then, in the second part, we will explore an alternative path for the same problem that makes use of a folding procedure.

The paper is organized as follows. 
In section~\ref{section2} we consider the specific case of a nonrelativistic particle on the circle $\mathbb{S}$, and 
explicitly show how one can generate boundary conditions by the action of the parity operator. In particular, we will show how to generate Dirichlet and Neumann boundary conditions by restriction of the Laplacian to suitable invariant subspaces which are eigenspaces of the parity operator. Then, in section~\ref{sec: genframe}, we will provide the general  framework for generating boundary conditions on a manifold without boundary via symmetry reduction.

In section~\ref{section3} we will show how to enlarge the class of possible boundary conditions  achievable by reduction. In particular, we will generate Robin boundary conditions starting from Neumann boundary conditions on a segment. This procedure will make use of a change in the metric of the interval. 

Then, in  section~\ref{section4},  we will  consider an alternative mechanism for generating boundary conditions, which makes use of a folding procedure.  It relies on unitary maps instead of projections, and can be applied to Hamiltonians that do not commute with the symmetry operator. 
The price to be paid is the introduction of a two-dimensional ancillary space, which physically represents an additional spin degree of freedom. We will consider the case of the momentum of a particle on a folded line and, in section~\ref{section5}, on a circle.
We finally draw our conclusions in section~\ref{section6}.

\section{Generation of boundary conditions by reduction}
\label{section2}

In this section we show a way of generating quantum boundary conditions by a reduction procedure.
We consider a compact manifold without boundary, and by modding out by a group action with fixed points we obtain a quotient manifold with boundaries.
Then, the projection of the quantum dynamics of a nonrelativistic particle on the initial manifold will give rise to a quantum dynamics on the quotient manifold, with specific quantum boundary conditions.

We first recall how to obtain a manifold with boundaries as a quotient of the action of a group of transformations on a manifold without boundaries. In particular we will focus on the unit circle  and the action of $\mathbb{Z}_2$ on it, which makes the unit circle collapse into an interval. 

The unit circle in the plane $\mathbb{R}^2$ is defined by
\begin{equation}
\mathbb{S}=\{(x_1,x_2)\in\mathbb{R}^2\,|\,x_1^2+x_2^2=1\},
\end{equation}
and can be parametrized, in $\mathbb{S}\setminus\{(-1,0)\}$, by
\begin{equation}
x\in(-\pi,\pi)\to
\begin{cases}
x_1=\cos x,  \\
x_2=\sin x.
\end{cases}
\end{equation}

It is possible to generate an interval of the real line by modding out the unit circle by a parity transformation. Consider the map
\begin{equation}
\label{eq:par}
\Pi:\mathbb{S}\to\mathbb{S}, \qquad
\Pi(x_1,x_2)=(x_1,-x_2),
\end{equation}
or in terms of $x\in (-\pi,\pi)$, $\Pi(x)=-x$.
Manifestly, $\Pi$ is a bijection and an involution, since $\Pi^2=\mathbb{I}$. 
\begin{figure}
\centering
\includegraphics[width=0.5\columnwidth]{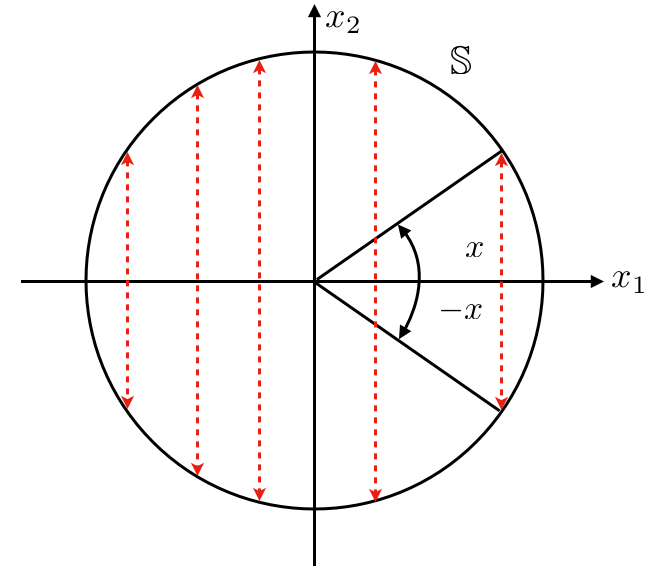}
\caption{Identification of symmetric points of the unit circle $\mathbb{S}$ by means of $\Pi$.}
\label{fig:circle}
\end{figure}

The action of $\Pi$ on the unit circle $\mathbb{S}$ (see Figure~\ref{fig:circle}) identifies pairs of points on the circle and admits only two fixed points, namely $(1,0)$ and $(-1,0)$.  
With the aid of the transformation $\Pi$ we are identifying symmetric points, or, equivalently, puncturing the circle in $(1,0)$ and $(-1,0)$, and pushing the lower semicircle onto the upper one. 

This is mathematically achieved by considering the quotient space of the unit circle under the action of the discrete group $\mathbb{Z}_2$. 
Indeed, the space of (discrete) orbits determined by $\Pi$,
\begin{equation}
M=\mathbb{S}/\Pi,
\end{equation}
is the interval $\mathbb{S}_+ =[0,\pi]$ (or, equivalently, the interval $\mathbb{S}_- =[-\pi,0]$). Thus, by taking the quotient of  the unit circle by the discrete action of $\Pi$ we obtain a one dimensional manifold with boundary, say $M=\mathbb{S}_+$.

Now, we will represent the action of $\Pi$ on square integrable functions on $\mathbb{S}$, and show how boundary conditions are going to emerge after this process. 

The action of $\Pi$ on functions can be implemented by a pull-back
\begin{equation}
P:L^2(\mathbb{S})\to L^2(\mathbb{S}),
\qquad (P\psi)(x)=\psi(\Pi (x))=\psi(-x).
\end{equation}
Moreover $P^2=\mathbb{I}$, so that the eigenspaces of the parity operator $P$ belong to the eigenvalues $\pm 1$.

The action of $P$ splits the Hilbert space $L^2(\mathbb{S})$ into two mutually orthogonal subspaces $\mathcal{H}_+$ and $\mathcal{H}_-$, defined by
\begin{equation}
\mathcal{H}_{\pm}=\{\psi\in L^2(\mathbb{S})\, |\, P\psi=\pm\psi\}.
\label{eq:evenodd}
\end{equation}
Notice that $L^2(\mathbb{S})$ can be identified with $L^2(-\pi,\pi)$, the Hilbert space of square integrable functions on the interval $(-\pi,\pi)$. Under such identification we get
\begin{equation}
\mathcal{H}_{\pm}= \{\psi\in L^2(-\pi,\pi)\,|\, \psi(-x)=\pm\psi(x)\},
\label{eq:Hpmdef}
\end{equation}
that is, the set of even and odd functions on $(-\pi,\pi)$, respectively.  

Consider now the Hamiltonian of a free particle on a circle~\eqref{eq:ham}. 
Since $\mathbb{S}$ is a compact manifold without boundary, the Laplace operator is essentially self-adjoint on $C^\infty(\mathbb{S})=C^\infty_c(\mathbb{S})$, the smooth function on the circle~\cite{jost}.
The domain of self-adjointness is the second Sobolev space $\mathrm{H}^2(\mathbb{S})$, which, in coordinates reads
\begin{equation}
\mathrm{H}^2(\mathbb{S})=\{\psi\in \mathrm{H}^2[-\pi,\pi]: \psi(-\pi)=\psi(\pi)\,,\,\psi'(-\pi)=\psi'(\pi)\}.
\label{eq:Sobolevdef}
\end{equation}
Here, $\mathrm{H}^2[-\pi,\pi]$ is the set of square-integrable functions, with square-integrable (first and) second distribution derivative.

Interestingly, the parity operator $P$ and the operator $H$ commute on $\mathrm{H}^2(\mathbb{S})$: 
\begin{equation}
H\,P=P\,H. 
\end{equation}
This is a crucial ingredient in our construction, which we remind consists in obtaining a self-adjoint operator on the quotient space, say the interval $[0,\pi]$, starting from the operator $H$ on the unit circle.

From the commutation relation $HP=PH$ it follows that whenever the operator $H$ acts on $\mathcal{H}_+$ (respectively on $\mathcal{H}_-$) then, its image remains in $\mathcal{H}_+$ (respectively in $\mathcal{H}_-$). Thus, the restriction of $H$ to one of the two subspaces gives rise to a self-adjoint operator.
We are going to show that the restrictions of $H$ to these parity eigenspaces can be identified with two self-adjoint Hamiltonian operators on the interval $[0,\pi]$.

From~\eqref{eq:Hpmdef} and~\eqref{eq:Sobolevdef}, one has
\begin{equation}
\label{eq:dom}
D(H|_{\mathcal{H}_+})=\mathrm{H}^2(\mathbb{S})\cap \mathcal{H}_+=\{\psi\in \mathrm{H}^2[-\pi,\pi]\cap \mathcal{H}_+:\psi'(-\pi)=0=\psi'(\pi)\}.
\end{equation}
Since the space of square integrable even functions $\psi$ on the interval $(-\pi,\pi)$ is unitarily equivalent to the space of square integrable functions $\phi$ on $(0,\pi)$, the domain in equation~\eqref{eq:dom} can be recast on the interval $\mathbb{S}_+=[0,\pi]$. Indeed, let us define the following unitary operator
\begin{eqnarray}
U_+: \mathcal{H}_+\to L^2(0,\pi), & & \phi(x)=(U_+\psi)(x)= \sqrt{2}\,\psi(x), \quad x\in \mathbb{S}_+,
\nonumber\\
U^\dagger_+: L^2(0,\pi)\to \mathcal{H}_+ , && \psi(x) = (U^\dagger_+\phi)(x)= \frac{1}{\sqrt{2}}\begin{cases}
  \phi(x),    & x \in \mathbb{S}_+ , \\
 \phi(-x) ,    & x \in \mathbb{S}_-.
\end{cases}
\end{eqnarray}
Then we have
\begin{eqnarray}
H_+:= U_+\,H|_{\mathcal{H}_+}\,U^\dagger_+= -\frac{\hbar^2}{2 m} \frac{{\d}^2}{\d x^2},
\end{eqnarray}
\begin{equation}
\label{eq:neum}
D(H_+)=\{\psi\in \mathrm{H}^2[0,\pi]:\psi'(0)=0=\psi'(\pi)\},
\end{equation}
where the derivative at 0 must vanish, because even functions have odd derivatives.
Equation~\eqref{eq:neum} can be immediately read on the quotient space $\mathbb{S}_+=[0,\pi]$, as a self-adjoint extension of the Hamiltonian describing a free particle on the interval $[0,\pi]$ with \emph{Neumann} boundary conditions.

Similarly, for the subspace of odd functions $\mathcal{H}_-$, we get
\begin{equation}
D(H|_{\mathcal{H}_-})=\mathrm{H}^2(\mathbb{S})\cap \mathcal{H}_-=\{\psi\in \mathrm{H}^2[-\pi,\pi]\cap \mathcal{H}_-:\psi(-\pi)=0=\psi(\pi)\},
\end{equation}
and we can define the unitary operator between the space of square integrable odd functions  $\psi$ on $(-\pi,\pi)$ and the space of square integrable functions $\phi$ on $(0,\pi)$ acting as
\begin{eqnarray}
U_-: \mathcal{H}_-\to L^2(0,\pi), && \phi(x) = (U_-\psi)(x)= \sqrt{2}\,\psi(x), \quad x \in \mathbb{S}_+ ,\nonumber\\
U^\dagger_-: L^2(0,\pi)\to \mathcal{H}_-,  && \psi(x)=(U^\dagger_-\phi)(x)= \frac{1}{\sqrt{2}}\begin{cases}
  \phi(x),   & x \in \mathbb{S}_+, \\
 -\phi(-x),     & x \in \mathbb{S}_-.
\end{cases}
\end{eqnarray}
Then, the restricted operator can be unitarily mapped into
\begin{eqnarray}
H_-= U_-\,H|_{\mathcal{H}_-}\,U^\dagger_-= -\frac{\hbar^2}{2 m} \frac{\d^2}{\d x^2},
\end{eqnarray}
\begin{equation}
D(H_-)=\{\psi\in \mathrm{H}^2[0,\pi]:\psi(0)=0=\psi(\pi)\}.
\end{equation}
In this case we have obtained a free particle on an interval with \emph{Dirichlet} boundary conditions.

Summing up, we started from a self-adjoint operator $H$ on the unit circle $\mathbb{S}$, which generates a unitary dynamics for the free particle on the circle. Besides, we picked out the eigenspaces of the parity $P$, say $\mathcal{H}_+$ and $\mathcal{H}_-$, which are left invariant by the one-parameter unitary group generated by $H$, since $[H,P]=0$.  Therefore, the operator $H$ restricted to the invariant subspaces $\mathcal{H}_{\pm}$, is still self-adjoint. Finally, the operators restricted to $\mathcal{H}_+$ and $\mathcal{H}_-$ can be read as two different self-adjoint realizations of the one-dimensional free Hamiltonian~\eqref{eq:ham} on the interval~$[0,\pi]$, with different boundary conditions. 

\section{The general framework}
\label{sec: genframe}
In the previous section we have shown how to generate boundary conditions on an interval starting from a unitary dynamics on the circle. In this section we would like to provide the reader with the general construction.

Consider a finite dimensional complex vector bundle $E\to M$ on a manifold $M$ carrying a Hermitian product. In the following we are going to denote the typical fiber by $V$ and the space of square integrable sections of $E$ by $L^2(M,V)$. Moreover, we suppose that the bundle is parallelizable.
Consider a set of fiberwise maximal pairwise commuting operators acting as a discrete group $G$ on $M$.

We denote by $\tilde M =M/G$ the orbifold obtained in the quotient process which can happen to be a manifold with boundary or with corners. The maximality of the set implies that $E$ admits a vector field of eigenvectors of this maximal set of operators. We obtain several copies of $L^2(\tilde M,\mathbb{C})$ , when we consider the Hilbert space of square integrable sections with values in a given joint eigenspace. 

Next, consider a one-parameter group of unitary bundle automorphisms on $E$, say $U(t): E\to E$, such that $G\,U(t)=U(t)\,G$. The latter condition implies that the one-parameter unitary group leaves unchanged every single copy of $L^2(\tilde M,\mathbb{C})$ and its infinitesimal generator still remains self-adjoint on the Hilbert space $L^2(\tilde M,\mathbb{C})$ associated to a given eigenvector. In general, with each eigenvector (one dimensional eigenspace), we obtain a different self-adjoint generator.
\subsection{Example 1}
Let $M$ be a compact Riemannian manifold without boundary. The Laplace-Beltrami operator is essentially self-adjoint, therefore its closure will generate a one-parameter group of unitary transformations on any complex vector bundle on $M$, with infinitesimal action $\Delta \otimes \mathbb{I}_n$ on the sections, where $\mathbb{I}_n$ is the identity matrix on $\mathbb{C}^n$. We can consider a discrete group acting on $E$ in terms of unitary transformations and extract from it a maximal set of fiberwise commuting operators.

In this manner we get a decomposition of the fiber $V$ into one-dimensional vector spaces and therefore $L^2(M,V)$ will be a direct sum of complex-valued square integrable functions. We select a basis of the complex-vector bundle, which is assumed to be parallelizable. With a unitary transformation it is always possible to consider a basis of eigenvectors of the commuting elements of the discrete group $G$. 

If the action commutes with $\Delta \otimes \mathbb{I}_n$ we return to the general arguments. As our operator is $\Delta \otimes \mathbb{I}_n$ it is clear that we only need our operator to commute with the action of the discrete group on $M$ so that it will be projectable onto~$\tilde M$.

We should notice that, while $\Delta$ will be in the enveloping algebra of first order differential operators acting on $M$, say vector fields acting on $M$, the same property will not hold true on $\tilde M$, because the projected Laplacian $\tilde\Delta$, does not need to be in the enveloping algebra of the derivations of $\mathcal{F}(\tilde M)$.

For example, let $M=\mathbb{S}^2=\{(x_1,x_2,x_3)\in\mathbb{R}^3\,|\,x_1^2+x_2^2+x_3^2=1\}$ and consider $\Delta=J^2_x+J^2_y+J^2_z$ and $\Pi:(x_1,x_2,x_3)\in\mathbb{R}^3\to(x_1,x_2,-x_3)$. The quotient space will be a disk, a manifold with a smooth boundary. The operator $J_z$ will pass to the quotient but $J_x$ and $J_y$ will not.

As a second example, consider again the free particle on a circle. Then $M= \mathbb{S}$, the Laplace operator $\Delta= \partial^2_{x_1}+\partial^2_{x_2}$ and the parity transformation $\Pi:(x_1,x_2)\to(x_1,-x_2)$. Let us denote by $x$ a coordinate of $M/\Pi$, and with $\tilde\Delta$ the Laplace operator on $M/\Pi$, say  $\tilde\Delta=\partial^2_x$. The only complete vector field will be $(x-1)(x+1)\partial_x$, therefore we have to investigate the domains of self-adjointness without being able to rely on the Lie algebra of complete vector fields acting on the quotient. 

\subsection{Example 2}
Let us consider the case of a spin-1/2 particle on the unit circle. In this case we have to consider the bundle $\mathbb{S}\times\mathbb{C}^2\to\mathbb{S}$, and apply the former construction to sections of $L^2(\mathbb{S},\mathbb{C}^2)$. 

We consider the operator $P:\psi(x_1,x_2)\to (n\cdot\sigma)\psi(x_1,-x_2)$, where $n$ is a unit vector in $\mathbb{R}^3$ and $\sigma=(\sigma_x,\sigma_y,\sigma_z)$, the vector of the three Pauli matrices. For the sake of simplicity we can consider the unit vector $n=(0,0,1)$, so that the operator $P$ reads $P\psi(x_1,x_2)=\sigma_3\psi(x_1,-x_2)$.  

Since $P^2=I$, the operator $P$ admits only two eigenvalues, say $\pm1$, which split the Hilbert space $L^2(\mathbb{S},\mathbb{C}^2)$ into $\mathcal{K}_{+}\oplus\mathcal{K}_{-}$, where $\mathcal{K}_{\pm}$ are the eigenspaces of $P$ with eigenvalues $\pm 1$.
More explicitly: 
\begin{equation}
\mathcal{K}_+=\left\{\psi=\begin{pmatrix}
\psi_1\\
\psi_2\\
\end{pmatrix}\in L^2(\mathbb{S},\mathbb{C}^2): \psi_1\in \mathcal{H}_+, \psi_2 \in \mathcal{H}_-\right\} ,
\end{equation}
\begin{equation}
\mathcal{K}_-=\left\{\psi=\begin{pmatrix}
\psi_1\\
\psi_2\\
\end{pmatrix}\in L^2(\mathbb{S},\mathbb{C}^2): \psi_1\in \mathcal{H}_-, \psi_2 \in \mathcal{H}_+ \right\} ,
\end{equation}
where $\mathcal{H}_+$ and $\mathcal{H}_-$ are, respectively, the space of even and odd functions on the circle as given in equation~\eqref{eq:Hpmdef}. Thus $\mathcal{K}_+=\mathcal{H}_+\oplus \mathcal{H}_-$ and since the Laplace operator, $\Delta \otimes \mathbb{I}$,  commutes with the operator $P$, the dynamics can be projected on $\tilde M$, the segment, and once again we can find Dirichlet or Neumann boundary conditions. In the same manner we obtain the same boundary conditions working with $\mathcal{K}_-$

To get additional extensions we might consider the fiber bundle $\mathbb{S}\times\mathbb{C}^4\to\mathbb{S}$, where the parity transformations may be implemented by $\{\sigma_3\otimes\sigma_3,\sigma_0\otimes\sigma_0,\sigma_0\otimes\sigma_3,\sigma_3\otimes\sigma_0\}$. We could use $G=\{\sigma_3\otimes\sigma_0, \sigma_0\otimes\mathfrak{u}(2)\}$ with maximal pairwise commuting operators $\{\sigma_3\otimes\sigma_0,\sigma_0\otimes n\cdot \sigma,\sigma_0\otimes\sigma_0,\sigma_3\otimes n\cdot\sigma\}$.
In this manner we should obtain additional self-adjoint extensions of the Laplace operator on the interval.

\section{General boundary conditions}
\label{section3}
In section~\ref{section2} we showed how to obtain Dirichlet and Neumann boundary conditions by a reduction of the free dynamics on the circle. Now, we would like  to get general  boundary conditions. We can move from functions on $\mathbb{S}$ to sections and consider covariant derivatives instead of ordinary ones. Any section over the circle can be trivialized at the cost of bringing in a connection and replacing ordinary derivatives with covariant ones. 

We are thus considering a U$(1)$ principal fiber bundle.
We write:
\begin{equation}
\label{eq:connect}
A= i \alpha(x) \mathrm{d}x,
\end{equation}
\begin{equation}
D\psi=\mathrm{d}\psi + A\psi,
\end{equation}
where $\mathrm{d}$ is the exterior derivative. We must ensure that  the connections are projectable under the map $P$:
\begin{equation}
P\,D=D\,P.
\end{equation}
Applying this expression on a section we get
\begin{equation}
\mathrm{d}\psi(-x)+A(-x)\psi(-x)=\mathrm{d}\psi(-x) + A(x)\psi(-x),
\end{equation}
which implies that
\begin{equation}
\alpha(x)=-\alpha(-x),
\end{equation}
and $\alpha(0)=0$. Thus $\alpha$ is an odd function and vanishes at the boundary.

Thus, if we restrict to even and odd subspaces we obtain
\begin{equation}
P\psi=-\psi\qquad\qquad \psi(0)=0=\psi(\pi),
\end{equation}
or
\begin{equation}
P\psi=\psi\qquad\qquad (D\psi)(0)=0=(D\psi)(\pi),
\end{equation}
and we can only get back Neumann or Dirichlet boundary conditions. Since $\alpha(0)=0$ we do not even get mixed Neumann or Dirichlet boundary conditions.

Let us make some observations which will help us solve the problem. Self-adjointness has to do with conservation of probability. Local boundary conditions assert that the probability current leaving the system at each boundary point vanishes:
\begin{equation}
j=-i(\bar\psi\nabla\psi-\psi\nabla\bar\psi)\qquad j(0)=0\,,\,j(\pi)=0,
\end{equation}
while, for non-local boundary conditions the current leaving one boundary point can be compensated by the one entering from the other side:
\begin{equation}
j(0)-j(\pi)=0,
\end{equation}
where the minus sign reflects the reversed orientation of the current.
Moreover, non-local boundary conditions cannot be obtained from parity reduction, since currents are odd under parity transformations:
\begin{equation}
P\,j\,P= -\,j,
\end{equation}
and as such the current at each boundary point is bound to vanish.

In order to get non-local boundary conditions we have to lift the action to the fiber and consider the combination of charge and parity transformation, say $C P$, rather than $P$ solely:
\begin{equation}
(C P \psi )\, (x)=\bar\psi(-x),
\end{equation}
Indeed, $C P$ acts not only on the base manifold but also on the U$(1)$ fiber and can reverse the orientation on both of them. The net effect is that $j$ is even under $C P$, namely
\begin{equation}
(CP)\,\,j\,(CP)= \,j,
\end{equation}
so that we can have non-vanishing currents at each boundary point.
From the former equation we can infer that non-local boundary conditions can emerge as a consequence of charge-parity transformations.

Within local boundary conditions we are going to prove that Robin boundary conditions~\eqref{eq:robin} can be generated by means of parity reduction, as long as we consider the Levi-Civita connection rather than a gauge connection (compare with equation~\eqref{eq:connect}).

We will prove that by changing the metric only in a small boundary layer we can get Robin boundary conditions starting from Neumann boundary conditions.
This construction can be physically realized for a vibrating string by introducing a localized non uniformity at the ends of the string.

The relevant quantities in  our problem are the spatial metric,
\begin{equation}
\mathrm{d}s^2=\d x^2 ,
\end{equation} 
and the Hamiltonian,
\begin{equation}
H=-\frac{\hbar^2}{2m}\frac{\d^2}{\d x^2},
\end{equation}
defined on $D(H_+)= \{\psi\in \mathrm{H}^2[0,\pi]: \psi'(0)=0=\psi'(\pi)\}$ with Neumann boundary conditions. 

Consider the following change of coordinates on the interval $[0,\pi]$:
\begin{equation}
\label{eq:chcor}
x\mapsto y=F(x)\qquad y=F(x)=\int_{0}^x f(t)\mathrm{d}t,
\end{equation}
where $f$ is a positive function on $[0,\pi]$, such that $\int_{0}^{\pi}f(t)\mathrm{d}t=\pi$. It is easy to see that this change of coordinates leaves the endpoints of the interval unchanged, while the metric reads
\begin{equation}
\mathrm{d}s^2=\biggl(\frac{\mathrm{d}x}{\mathrm{d}y}\biggr)^2\mathrm{d}y^2=\frac{1}{[f(y)]^{2}}\mathrm{d}y^2 .
\end{equation}
The new wavefunction $\phi$ changes according to the unitary transformation:
\begin{equation}
U_f: L^2((0,\pi),\mathrm{d}x)\to L^2((0,\pi),\mathrm{d}y),
\end{equation}
\begin{equation}
\label{eq:uni}
\phi(y)=(U_f\,\psi)(y)=\frac{1}{\sqrt{g(y)}}\psi(F^{-1}(y)), \qquad g(y)=f(F^{-1}(y)),
\end{equation}
because, from a local point of view, a local change of coordinates cannot change the probability: $|\psi|^2\mathrm{d}\,x=|\phi|^2\mathrm{d}\,y$.
Under this unitary transformation, the momentum operator $p=-i\hbar\,\d/\d x$ becomes
\begin{equation}
\label{eq:newmom}
p_f =U_f\,p\,U^\dag_f =g p-\frac{i\hbar }{2} g'.
\end{equation}

Accordingly, the transformed Hamiltonian reads
\begin{equation}
\label{eq:newham}
H_f  =U_f\,H\,U^\dag_f  =- g^2\frac{\hbar^2}{2m} \frac{\d^2}{\d y^2} -g g'  \frac{\hbar^2}{m} \frac{\d}{\d y} + V ,
\end{equation}
where
\begin{equation}
V=\frac{\hbar^2}{8m}\left[(g')^2+2 g g'' \right] .
\end{equation}

Next, we would like to understand how the Neumann boundary conditions change under this coordinate transformation. 
In order to do so we compute the first derivative of $\psi(x)=\sqrt{f(x)}\phi(y(x))$:

\begin{equation}
\psi'(x)=\frac{1}{2\sqrt{f(x)}}f'(x)\,\phi(F(x)) + f(x)\sqrt{f(x)}\phi'(F(x)).
\end{equation}

Then, at the boundary, where the functions have to vanish, we find that
\begin{equation}
\phi'(F(x))=-\frac{1}{2[f(x)]^2}f'(x)\phi(F(x)),
\end{equation}
that is to say
\begin{equation}
\begin{cases}
\phi'(0)=\nu_0\,\phi(0)      ,\\
\phi'(\pi)=-\nu_{\pi}\,\phi(\pi)       ,
\end{cases}
\end{equation}
where $\nu_0=-\frac{1}{2}\frac{f'(0)}{[f(0)]^2}$, $\nu_{\pi}=\frac{1}{2}\frac{f'(\pi)}{[f(\pi)]^2}$ and where we used the relations: $y(0)=0$ and $y(\pi)=\pi$.
\begin{figure}
\centering
\label{fig:funct}
\includegraphics[width=0.6\columnwidth]{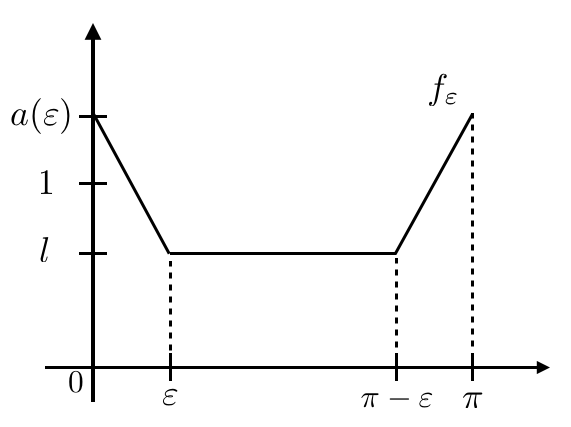}
\caption{Plot of the functions $f_\varepsilon$, defined in\eqref{eq:feps}, which are used for generating Robin boundary conditions.}
\end{figure}

By a change of coordinates, as in equation~\eqref{eq:chcor}, we managed to induce Robin boundary conditions starting from Neumann boundary conditions. However, also the original physical problem---a free quantum particle in a one-dimensional box---was changed, since, after the transformation in equation~\eqref{eq:uni} we obtained a new Hamiltonian~\eqref{eq:newham} with a varying mass and a potential energy term $V(y)$.

In order to overcome this drawback, we will consider a sequence of functions $f_\varepsilon(x)$, $\varepsilon>0$, which tends to a constant function in the limit $\varepsilon\to 0$, namely 
$f_\varepsilon(x)\to k$,
pointwise for all $x\in(0,\pi)$. In principle $k$ may even diverge, but, as we are going to see in the following, this would represent an unphysical situation.
With this assumption the Hamiltonian $H_{f_\varepsilon}$ in equation~\eqref{eq:newham}, converges in the bulk to the free particle Hamiltonian, with a renormalized  mass $M=m/k^2$, that is
\begin{equation}
H_{f_\varepsilon}\to -\frac{\hbar^2}{2m}k^2\frac{\d^2}{\d y^2}=-\frac{\hbar^2}{2M}\frac{\d^2}{\d y^2}.
\end{equation}
Moreover we suppose that the following limits for the  $\varepsilon$-dependent  Robin constants exist:
\begin{eqnarray}
\lim_{\varepsilon\to0} \nu_0=\mu_0>0,\\
\lim_{\varepsilon\to0} \nu_\pi=\mu_\pi>0.
\end{eqnarray}
For example, consider the following change of coordinates as shown in figure~\ref{fig:funct}:
\begin{equation}
f_\varepsilon (x)=
\begin{cases}
\frac{(l-a)x}{\varepsilon}+a\qquad &0\le x\le\varepsilon,\\
l\qquad &\varepsilon\le x\le\pi-\varepsilon,\\
a-\frac{(l-a)(x-\pi)}{\varepsilon} &\pi-\varepsilon\le x\le\pi.
\end{cases}
\label{eq:feps}
\end{equation}
where $l=\frac{\pi-\varepsilon a}{\pi-\varepsilon}$ is such that $\int_0^\pi f_\varepsilon=\pi$ for every $\varepsilon>0$ and $a=a(\varepsilon)$ is a function of $\varepsilon$.
The Robin parameter at $x=0$, say $\nu_0$, reads
\begin{equation}
\nu_0=-\frac{1}{2}\frac{f'_\varepsilon(0)}{[f_\varepsilon(0)]^2}=\frac{\pi}{2a^2\varepsilon}\left(\frac{a-1}{\pi-\varepsilon}\right).
\end{equation}

If $a(\varepsilon)=1/2\mu_0\varepsilon$, then in the limit $\varepsilon\downarrow0$, the constant $\nu_0$ converges to the fixed parameter $\mu_0$. Interestingly, in the bulk $f_\varepsilon$ converges to the value $\frac{2\mu_0\pi-1}{2\mu_0\pi}$. Thus, in the interior of $(0,\pi)$, the Hamiltonian in equation~\eqref{eq:newham} converges to the Hamiltonian of a free particle, with a renormalized mass $M=m\left(\frac{2\mu_0\pi}{2\mu_0\pi-1}\right)^2$: 
\begin{equation}
H_{f_\varepsilon}\to -\frac{\hbar^2}{2m}\left(\frac{2\mu_0\pi-1}{2\mu_0\pi}\right)^2\frac{\d^2}{\d y^2}=-\frac{\hbar^2}{2M}\frac{\d^2}{\d y^2}.
\end{equation}

If  $a(\varepsilon)$, instead, diverges more slowly than $1/\varepsilon$ (and does not converge to $1$), then the constant $\nu_0$ converges to $0$, that is to say to a Dirichlet boundary condition at $0$. In this case, the limiting Hamiltonian in the bulk is that of a free particle with mass $m$
\begin{equation}
H_{f_\varepsilon}\to -\frac{\hbar^2}{2m}\frac{\d^2}{\d y^2},
\end{equation}
because the height $l$ converges to $1$ as $\varepsilon \downarrow 0$.

Finally, if $a(\varepsilon)\to1$, we find the Hamiltonian of a free particle with Neumann boundary conditions, as we could have expected from the very beginning. 
On the other hand, if $a(\varepsilon)$ diverges faster than $1/\varepsilon$ as $\varepsilon \downarrow 0$ we get an  unphysical limit. In this situation, indeed, the height $l$ diverges, which corresponds to a vanishing mass limit of the Hamiltonian in equation~\eqref{eq:newham}. 

So far, we have considered only what happens at $x=0$. Analogously one can discuss the case for the other endpoint of the interval, say $x=\pi$, getting the same results obtained for $x=0$, say $\mu_0=\mu_\pi$. In order to get different Robin parameters it is sufficient to consider at $x=\pi$ a value different from $a(\varepsilon)$, and then repeat the previous procedure.

\section{Generation of boundary conditions by folding}
\label{section4}
In the previous sections we have shown how to generate quantum boundary conditions by means of a quotient procedure on the base manifold. 
By taking the quotient of a manifold without boundary (e.g.\ the circle) with respect to the action of a finite group (e.g.\ $\mathbb{Z}_2$), we have obtained a manifold with boundary (e.g.\ the interval).
Then, we have considered  the $L^2$ space over the original manifold and taken a subspace (e.g.\ the space of even/odd wave functions)  which is invariant under the action of the Hamiltonian (e.g.\ the Laplacian) and can be identified with the $L^2$ space over the quotient manifold. Thus a projection of the original quantum dynamics onto that subspace has provided the quantum dynamics on the manifold with boundary, equipped with specific quantum boundary conditions (e.g.\ Neumann/Dirichlet).

In the following sections we are going to show how to generate quantum boundary conditions by means of a folding procedure. 
At variance with the previous strategy, here we will establish a unitary map, instead of a projection, between suitable $L^2$ spaces over the original and the folded base manifolds. We will show that the requirement of unitarity implies the emergence of an additional spin degree of freedom in the quantum dynamics on the manifold with boundary.

In this section we consider the folding of a line into a half-line, and in the following section we will consider again the case of a circle. 
As a starting operator we will always take the momentum operator, which does not have self-adjoint realizations on the half-line and on the interval (with local boundary conditions), and thus cannot generate unitary dynamics. We will show how the emerging spin degree of freedom will be of help to restore  unitarity.

Consider the momentum operator on the real line,
\begin{equation}
p=-i\hbar\frac{\mathrm{d}}{\mathrm{d}x},
\end{equation}
defined on its domain of self-adjointness,
\begin{equation}
D(p)=\mathrm{H}^1(\mathbb{R})= \{\psi\in L^2(\mathbb{R}) \,\, |\,\, \psi'\in L^2(\mathbb{R})\,\},
\end{equation}
where $\mathrm{H}^1(\mathbb{R})$ is the first Sobolev space, of square integrable functions with square-integrable distributional derivative.

Let $\mathbb{R}_+=\{x\in\mathbb{R}:x\ge0\}$ be the positive half-line. We are going to construct a natural unitary map between $L^2(\mathbb{R})$ and $L^2(\mathbb{R}_+)\otimes \mathbb{C}^2$. Next, we will use this  map to find out the operator on $L^2(\mathbb{R}_+)\otimes \mathbb{C}^2$ into which  the original momentum operator on $L^2(\mathbb{R})$ is transformed.
This procedure maps a self-adjoint operator in $L^2(\mathbb{R})$ into a self-adjoint operator in $L^2(\mathbb{R}_+)\otimes \mathbb{C}^2$. This fact is extremely interesting from a physical perspective, because, as mentioned above, the momentum operator admits no self-adjoint extensions on the  half-line, say on $L^2(\mathbb{R}_+)$, since there is a net probability flux through the boundary at the origin, which cannot be compensated~\cite{reed}.

The above procedure, nevertheless, will produce a self-adjoint momentum operator on the half-line at the price of the introduction of an ancillary space, $\mathbb{C}^2$. Such an operator can be physically interpreted as a Dirac operator for a spin-1/2 particle on the  half-line $\mathbb{R}_+$.

We define the  map
\begin{eqnarray}
\label{eq:udagger}
U: L^2(\mathbb{R})&\to & L^2(\mathbb{R}_+)\otimes \mathbb{C}^2,\nonumber\\
\psi(x)&\mapsto&
\Phi(y)= \begin{pmatrix}
\phi_+(y)\\
\phi_-(y)\\
\end{pmatrix} =  (U\psi)(y)=
\begin{pmatrix}
\psi(y)\\
\psi(-y)\\
\end{pmatrix}.
\end{eqnarray}
where $x \in \mathbb{R}$ and $y \in \mathbb{R}_+$.
Its adjoint reads
\begin{eqnarray}
\label{eq:udagger1}
U^\dagger: L^2(\mathbb{R}_+)\otimes \mathbb{C}^2\nonumber&\to&L^2(\mathbb{R}), \\
\Phi(y) = \begin{pmatrix}
\phi_+(y)\\
\phi_-(y)\\
\end{pmatrix}&\mapsto&
\psi(x) = (U^\dagger \Phi)(x) =
\begin{cases}
\phi_+(x) &\mathrm{if}\ x\in \mathbb{R}_+ \\
\phi_-(-x) &\mathrm{if}\ x\in \mathbb{R}_-\\
\end{cases} .
\end{eqnarray}
It can be easily verified that $U$ is unitary, namely  $UU^\dagger=U^\dagger U=\mathbb{I}$.

Since the wave functions $\psi$ in $D(p)= \mathrm{H}^1(\mathbb{R})$ are continuous, one has that $\psi(0^+)=\psi(0^-)$. Therefore, 
the domain of the transformed operator $\tilde p=U p U^\dagger$ is 
\begin{equation}
D(\tilde p )= U D(p) =\{ \Phi \in \mathrm{H}^1(\mathbb{R}_+)\otimes \mathbb{C}^2\,| \, \phi_+(0)=\phi_-(0)\,\}.
\end{equation}
It is clear from the above expression that a boundary condition has naturally emerged after this unitary transformation. 
\begin{figure}
\centering
\includegraphics[width=0.9\columnwidth]{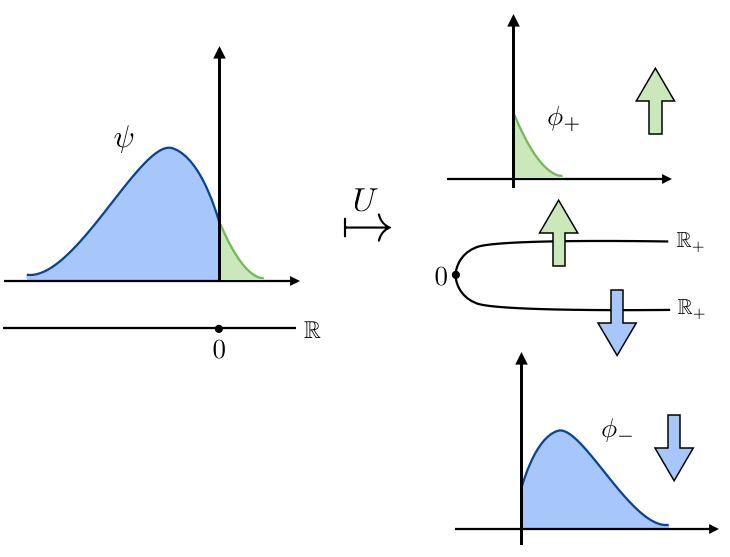}
\caption{Action of the unitary operator $U$ defined in~\eqref{eq:udagger}. The splitting of the wavefunction $\psi$ into the two spinorial components, $\phi_+$ and $\phi_-$, is represented.}
\label{fig:funct1}
\end{figure}

Let us now look at the explicit form of the operator $\tilde p=U p U^\dagger$. We get
\begin{eqnarray}
 (p U^\dagger \Phi) (x) = p U^\dagger
\begin{pmatrix}
\phi_+\\
\phi_-
\end{pmatrix} (x) &=& p \begin{cases}
\phi_+(x) &\mathrm{if}\ x\in \mathbb{R}_+ \\
\phi_-(-x) &\mathrm{if}\ x\in \mathbb{R}_-
\end{cases} 
\nonumber\\
&=& - i\hbar \begin{cases}
\phi'_+(x) &\mathrm{if}\ x\in \mathbb{R}_+ \\
-\phi'_-(-x) &\mathrm{if}\ x\in \mathbb{R}_-
\end{cases} ,
\end{eqnarray}
whence
\begin{eqnarray}
\tilde p \Phi(y)=
U p U^\dagger
\begin{pmatrix}
\phi_+(y)\\
\phi_-(y)
\end{pmatrix}
= - i\hbar
\begin{pmatrix}
\phi'_+(y)\\
-\phi'_-(y)
\end{pmatrix}.
\end{eqnarray}

Therefore,
\begin{equation}
\tilde p=-i\hbar\frac{\d}{\d y}\otimes \sigma_z\qquad D(\tilde p )= \{ \Phi \in \mathrm{H}^1(\mathbb{R}_+)\otimes \mathbb{C}^2\,| \Phi(0)=\sigma_x \Phi(0)\,\},
\label{eq:ptilde}
\end{equation}
where $\sigma_x$ and $\sigma_z$ are the first and the third Pauli matrix, respectively.

In other words, we started with the momentum operator $p$ of a quantum particle on the line $L^2(\mathbb{R})$. Then, we punctured the  line at the origin and folded it, resulting into two copies of $L^2(\mathbb{R}_+)$, that is $L^2(\mathbb{R}_+)\otimes\mathbb{C}^2$. 
See Fig.~\ref{fig:funct1}.
Next, we showed that the momentum on the real line transforms into a Dirac operator on the  half-line with a definite quantum boundary condition which makes it self-adjoint.  

It is instructive to look at the above procedure in the opposite way, which would represent a dilation process: Suppose we start with the momentum operator on the  half-line, i.e.\ in $L^2(\mathbb{R}_+)$, which admits no self-adjoint extensions, because its deficiency indices are different~\cite{reed}. 
Then, in the spirit of Naimark's dilation theorem~\cite{agla}, one can instead enlarge the Hilbert space and look at an extension of the problem one has started with, which is significantly different.
In other words, through a dilation procedure, we can get the operator $\tilde p$, which is a Naimark extension of the momentum on the  half-line and has a different physical interpretation, as the Dirac operator of a spin-1/2 particle on the  half-line.

 From a physical point of view the new operator $\tilde p$ could represent a spin-1/2 particle interacting with a wall, which flips both the momentum and  the spin of the particle, through the operator $\sigma_x$ in the boundary conditions~\eqref{eq:ptilde}, and thus preserves its helicity. An alternative interpretation is given by a spinless particle on the half-line which collides with a detector at the boundary. The detector has  two possible states and corresponds to the two-level system. When the particle hits the boundary, it will bounce with a corresponding flip of its momentum, and  the detector will click.

In this construction the self-adjointness of the resulting operator relies on the ancillary spin. 

Indeed, the dynamics on the space $L^2(\mathbb{R}_+)\otimes\mathbb{C}^2$ 
is unitary, but this cannot be the case on the spatial component $L^2(\mathbb{R}_+)$, since its generator, the momentum operator, is not self-adjoint  on the half-line. 
The momentum operator on the line is not projectable onto the half-line, and this results in the projected operator losing self-adjointness. 

This issue can be detected by considering the projection of the space $L^2(\mathbb{R}_+)\otimes\mathbb{C}^2$, which is unitarily equivalent to $L^2(\mathbb{R})$, onto its spatial component $L^2(\mathbb{R}_+)$.   This projection, obtained by tracing out the spin component $\mathbb{C}^2$,
maps separable pure states into pure states, while entangled states are mapped into  mixed states. Therefore, if the unitary dynamics on $L^2(\mathbb{R}_+)\otimes\mathbb{C}^2$  generates entanglement, its projection cannot be unitary. 
This establishes an interesting link between entanglement generation of a unitary evolution and the lack of self-adjointness of the projected generator.

That is just the case of the example under consideration. Indeed, suppose that the initial state of the system is
\begin{equation}
\phi \otimes\frac{|{\uparrow}\rangle+|{\downarrow}\rangle}{\sqrt{2}},
\end{equation}
where $\phi\in L^2(\mathbb{R}_+)$ is a normalized wave packet which vanishes in a neighbourhood of the origin $x=0$, and $\{|{\uparrow}\rangle,|{\downarrow}\rangle\}$ is the eigenbasis of $\sigma_z$.
Then the evolved state for sufficiently small times $t$ reads
\begin{equation}
e^{-i t p \otimes \sigma_z}\left(\phi(x) \otimes\frac{|{\uparrow}\rangle+|{\downarrow}\rangle}{\sqrt{2}}\right)=\phi(x-t) \otimes\frac{|{\uparrow}\rangle}{\sqrt{2}}+ \phi(x+t) \otimes\frac{|{\downarrow}\rangle}{\sqrt{2}},
\end{equation}
and the spatial degrees of freedom gets manifestly entangled with the spinorial ones for positive times.

\section{Momentum operator on the circle}

\label{section5}

In this section we would like to provide the reader with another example of the folding procedure. We are going to study the momentum of a  particle on a circle $\mathbb{S}$ and, as in the previous section, we will map this problem into a unitarily equivalent one. As a consequence, boundary conditions will be generated in the transformed system.

We recall the natural identifications:
\begin{equation}
L^2(\mathbb{S})=L^2(-\pi,\pi)=L^2(-\pi,0)\oplus L^2(0,\pi),
\label{eq:identif}
\end{equation}
that will turn out to be useful in the following discussion. 
Consider the momentum operator of a particle on a circle
\begin{equation}
p=-i\hbar\frac{\mathrm{d}}{\mathrm{d} x},
\qquad 
D(p)=\mathrm{H}^1(\mathbb{S})= \{\psi\in \mathrm{H}^1[-\pi,\pi]\,\, |\,\, \psi(-\pi)=\psi(\pi)\,\}.
\end{equation}

By using the identifications~\eqref{eq:identif} and the continuity of the functions in the first Sobolev space $\mathrm{H}^1$, the domain of $p$ can be rewritten as
\begin{equation}
D(p)= \{\psi\in \mathrm{H}^1[-\pi,0]\oplus \mathrm{H}^1[0,\pi]\,\, |\,\, \psi(0^-)=\psi(0^+)\,,\,\psi(-\pi)=\psi(\pi)\}.
\end{equation}

We are going to unitarily map this problem on $L^2(0,\pi)\otimes\mathbb{C}^2$. 
\begin{figure}%[tbp]
\centering
\includegraphics[width=1\columnwidth]{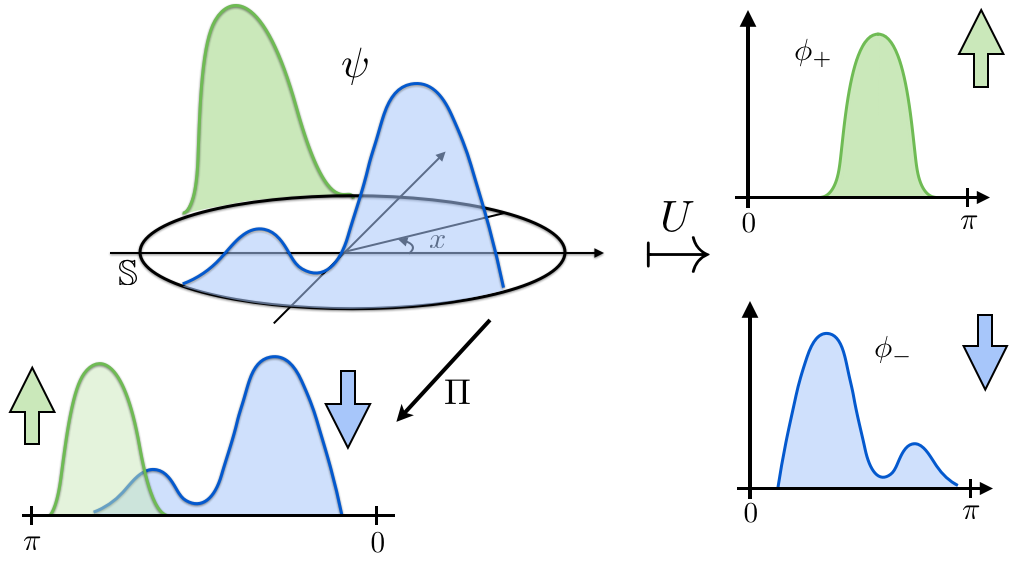}
\caption{Action of the unitary operator $U$ on $L^2(\mathbb{S})$ defined in~\eqref{eq:unicirc}.  The splitting of the wavefunction $\psi$ into the two spinorial components, $\phi_+$ and $\phi_-$, is represented (on the right). On the left the folding procedure on the interval $[0,\pi]$ is pictorially shown.}
\label{fig:unitcircle}
\end{figure}
Indeed the following map is unitary, as pictorially shown in figure~\ref{fig:unitcircle}:
\begin{eqnarray}
U: L^2(\mathbb{S})&\to&  L^2(0,\pi)\otimes\mathbb{C}^2, \\
\psi(x)&\mapsto&(U\psi)(y)=
\begin{pmatrix}
\phi_+(y)\\
\phi_-(y)
\end{pmatrix} = 
\begin{pmatrix}
\psi(y)\\
\psi(-y)
\end{pmatrix},
\label{eq:unicirc}
\end{eqnarray}
where $x \in [-\pi,\pi]$ and $y \in [0,\pi]$.
Its inverse reads
\begin{eqnarray}
U^\dagger: L^2(0,\pi)\otimes\mathbb{C}^2\nonumber&\to&L^2(\mathbb{S}),\\
U^\dagger
\begin{pmatrix}
\phi_+(y)\\
\phi_-(y)\\
\end{pmatrix} &= &
\begin{cases}
\phi_+(x) &\mathrm{if}\ x\in [0,\pi] \\
\phi_-(-x) &\mathrm{if}\ x\in [-\pi,0]
\end{cases} .
\end{eqnarray} 
The domain of the transformed operator $\tilde p=U p U^\dagger$ is 
\begin{equation}
D(\tilde p )= U D(p) 
=\{ \Phi \in \mathrm{H}^1[0,\pi]\otimes \mathbb{C}^2\,| \, \phi_+(0)=\phi_-(0)\,,\,\phi_+(\pi)=\phi_-(\pi)\}.
\end{equation}

and $\tilde{p}$ acts as
\begin{eqnarray}
\tilde p
\begin{pmatrix}
\phi_+(y)\\
\phi_-(y)\\
\end{pmatrix}=\,-i \hbar
\begin{pmatrix}
\phi'_+(y)\\
-\phi'_-(y)
\end{pmatrix} .
\end{eqnarray}
Therefore, we get
\begin{equation}
\tilde p= -i\frac{\d}{\d y}\otimes \sigma_z, \quad D(\tilde p )= \{ \Phi \in \mathrm{H}^1[0,\pi]\otimes \mathbb{C}^2\,| \Phi(0)=\sigma_x\,\Phi(0)\,,\,\Phi(\pi)=\sigma_x\,\Phi(\pi)\} .
\end{equation}
In a nutshell, we started from the momentum operator on the unit circle and by means of a unitary transformation we ended up with the Dirac operator on a segment with well-prescribed boundary conditions.

As in the previous example, we managed to obtain a spin-1/2 particle on a manifold with boundary starting from a spinless particle on a manifold without boundary.
Again, the emergent spin degrees of freedom are crucial in the conservation of probability, since the quantum boundary conditions imply both a spin flip  and a momentum flip  whenever the particle bounces off the boundary.

\section{Conclusions and outlook}

\label{section6}

We have considered the emergence of quantum boundary conditions when boundaries are generated  by  modding out a closed manifold by a group action with fixed points. In particular, we looked at  a free particle on the circle and at the action of the parity operator, and showed how a set of local quantum boundary conditions are generated. Moreover, we have showed that parity reduction cannot lead to non-local boundary conditions, and
no boundary conditions more general than Robin can be produced.

Therefore, one has to move from functions on the circle to sections and consider covariant derivatives instead of ordinary ones.
We have argued that, by lifting the action to the fibers and 
by making use of both  $C$ and $P$ one can
generate non-local boundary conditions, because the probability current is even under $CP$ and thus can be nonzero at the boundary points.
This is an interesting  mechanism which deserves to be investigated in detail.

Finally, we have exhibited an alternative procedure of generation of quantum boundary conditions by folding, which causes the emergence of an auxiliary spin Hilbert space. The additional degree of freedom is ancillary to the preservation of unitarity, and as such allows to consider also Hamiltonians which are not projectable. Indeed, we have shown how a unitary evolution on the manifold with boundaries is  provided by the entanglement between the spatial and the spinorial degrees of freedom. Further investigation will  be devoted to this unexpected link between self-adjoint extensions of symmetric operators and purifications of mixed states.

\section*{Acknowledgments}
We thank Sara Di Martino for  initial discussions on these topics.
This work was partially supported by the Italian National Group of Mathematical Physics (GNFM-INdAM), and by INFN through the project ``QUANTUM''.

\end{document}